# Doped graphene/carbon black hybrid catalyst giving enhanced oxygen reduction reaction activity with high resistance to corrosion in proton exchange membrane fuel cells

Zhaoqi Ji [1,2], Jianuo Chen [2], María Pérez-Page [2*], Zunmin Guo [2], Ziyu Zhao [2], Rongsheng Cai [3], Maxwell T. P. Rigby [3], Sarah J. Haigh [3], Stuart M. Homes [2*].

[1] School of Automotive Engineering, Harbin Institute of Technology, Weihai, 264209, Shandong, China

[2] Department of Chemical Engineering and Analytical Science, The University of Manchester, Manchester, M13 9PL, U.K.

[3] Department of Materials, The University of Manchester, Manchester, M13 9PL, U.K.

* Corresponding authors: stuart.holmes@manchester.ac.uk (S. M. Holmes); maría.pérez-page@manchester.ac.uk (M. Pérez-Page)

**Abstract**

Nitrogen doping of the carbon is an important method to improve the performance and durability of catalysts for proton exchange membrane fuel cells by platinum-nitrogen and carbon-nitrogen bonds. This study shows that *p*-phenyl groups and graphitic N acting bridges linking platinum and the graphene/carbon black (the ratio graphene/carbon black=2/3) hybrid support materials achieved the average size of platinum nanoparticles with (4.88 ± 1.79) nm. It improved the performance of the lower-temperature hydrogen fuel cell up to 0.934 W cm$^{-2}$ at 0.60 V, which is 1.55 times greater than that of commercial Pt/C. Doping also enhanced the interaction between Pt and the support materials, and the resistance to corrosion, thus improving the durability of the low-temperature hydrogen fuel cell with a much lower decay of 10 mV at 0.80 A cm$^{-2}$ after 30k cycles of an in-situ accelerated stress test of catalyst degradation than that of 92 mV in Pt/C, which achieves the target of Department of Energy (<30 mV). Meanwhile, Pt/NrEGO$_2$-CB$_3$ remains 78% of initial power density at 1.5 A cm$^{-2}$ after 5k cycles of in-situ accelerated stress test of carbon corrosion, which is more stable than the power density of commercial Pt/C, keeping only 54% after accelerated stress test.

degradation, Carbon corrosion.

## 1. Introduction

At present, there is an urgent need to find renewable energy alternatives to solve the energy depletion and environmental pollution caused by the massive extraction and use of fossil fuels. The hydrogen fuel cell is regarded as one of the most promising alternative renewable energy conversion technologies because of its higher energy efficiency, higher power density and zero-carbon emissions. Hydrogen oxidation reaction (HOR) and oxygen reduction reaction (ORR) underpin the performance of the hydrogen fuel cell. The reactions are commonly enhanced by different precious metal catalysts, which account for 41% of the total cost in a fuel cell stack [1]. Although, Pt and Pt alloy catalysts have made great progress in improving ORR activity with lower cost, the high Pt catalyst loading and poor durability of the hydrogen fuel cell remains a barrier to practical operation and commercial uptake [2]. The degradation of Pt catalysts is mainly caused by two mechanisms: i) Pt coalescence and agglomeration by Ostwald ripening and migration on the surface of support [2–4]; ii) carbon corrosion, which is a cause of Pt nanoparticles detaching from the carbon support, thus resulting in a loss of electrochemical surface area (ECSA) [5,6]. Therefore, the degradation of catalysts depends on the quality of the interaction between the Pt nanoparticles and the supports. Currently, the most common supporting materials are a kind of high surface area carbon, such as Vulcan XC-72, carbon black (CB), Ketjen black and carbon nanotubes [7–10]. Xie et al reported that a higher degree of graphitic structure facilitates higher carbon corrosion resistance [11]. As a result, novel graphitized materials, e.g. graphitic carbon nitride [12], graphitic mesoporous carbon [13], have been studied for their ability to overcome the carbon corrosion.

Reduced graphene oxide (rGO) has attracted particular attention as an ideal platinum nanoparticles (PtNPs) catalyst support material due to its large surface area, excellent electrical conductivity, good mechanical and electrochemical stability [14,15] and ease of manufacture (when compared to graphene). However, the difficulty of transporting the reactant gases ($H_2$ and $O_2$) to approach the active sites is a critical problem because of the tendency for the restacking of graphene flakes through π-π interaction [16]. To overcome this issue, rGO flakes were successfully intercalated with CB as a kind of hybrid support material [17]. Generally, rGO flakes obtained by the Hummers method face many problems such as environmental and safety issues associated with the reaction process as well as this being a high cost and time consuming preparation route [18]. In addition, more defects (higher oxygen functional groups) caused by incomplete reduction have a big influence on the electron transfer rate [19]. Therefore, the electrochemical exfoliation of graphite would be a good alternative to obtain exfoliated graphene oxide (EGO), which is faster (less than an hour to synthesis) than Hummers method, more environmentally friendly (no use of strong acids and oxidants) and has a tens of grams scale, showing potential for industrial production [20]. However, the lower activation energy of Pt atoms migrating between bridge sites on graphene results in facile movement of Pt atoms across graphene sheets [21]. Therefore, the Pt agglomerates during application in a fuel cell, resulting in poor durability.

To further enhance ORR activity and the durability of the catalyst, nitrogen has been widely introduced into support materials. The strong electron donor behavior of nitrogen (N) can enhance $\pi$ bonding and add a delocalized electron to create a desirable binding site and promote electron donation from graphite to oxygen, leading to an enhancement of the catalyst in terms of metal degradation as well as carbon corrosion [22–24]. Moreover, it has also been suggested that N could facilitate the ORR activity by polarizing the adjacent carbon [25]. N species mainly include pyridinic, $p$-phenyl –$NH_2$, pyrrolic N, quaternary N, graphitic N and N-oxides [26]. Pyridinic N and graphitic N have been widely introduced into graphene flakes to increase the defects as well as to provide electro-negativity to facilitate a uniform Pt atoms distribution, thus promoting ORR activity. However, the high amount of pyridinic N and graphitic N was always obtained by high-temperature annealing (> 900 °C), which is difficult to commercialize [27]. Electro-negative –$NH_2$, could also form either an ionic bond or a covalent bond with the Pt complex ions (i.e., $PtCl_6^{2-}$) [28]. However, to the best of our knowledge, there has been no study of –$NH_2$ and graphitic N doped reduced EGO (NrEGO) and CB as a hybrid supporting material with performance and durability testing in-situ in a hydrogen fuel cell.

This work reports the synthesis of –$NH_2$ groups and graphitic N doped reduced, electrochemically exfoliated, graphene oxide/carbon black hybrid supporting materials (NrEGO-CB) and investigates the effect of nitrogen doping to obtain better performance and durability for the hydrogen fuel cell. The in-situ accelerated stress test results suggest that NrEGO-CB materials have enhanced ORR activity through controlling the Pt nanoparticles size as well as resistance to carbon corrosion with higher electrochemical stability in the hydrogen fuel cell. Thus, they have potential as an advanced electrocatalyst support to further develop the performance and durability of the hydrogen fuel cell.

## 2. Experimental

*2.1 Materials*

The materials used were urea (99.00%, Sigma Aldrich), ethylene glycol (EG, >99.00%, Fisher Scientific), chloroplatinic acid hexahydrate ($H_2PtCl_6·6H_2O$, ACS reagent, >37.50% Pt basis, Honeywell Fluka), concentrated sulfuric acid (>95.00 % $H_2SO_4$, VWR Corporation), ammonium nitrate ($NH_4NO_3$, >98.50%, VWR Corporati), 60 wt% platinum supported on high surface area carbon catalyst (Pt/C, Alfa Aesar), Toray carbon paper (TGPH 090, 280 micron, Fuel Cell Store), Ketjen black (EC-300J, AkzoNobel), Polytetrafluoroethylene (60 wt% PTFE, Sigma Aldrich), 5 wt% and 20 wt% Nafion ionomer solution (Ion Power), iso-propanol (IPA, 99.50%, Fisher Scientific), acetone (99.60%, Fisher Scientific) and ethanol absolute (99.98%, Fisher Scientific). Electrochemically exfoliated graphene oxide (EGO) was prepared by a two-step electrochemical exfoliation method with concentrated sulfuric acid ($H_2SO_4$) and ammonium nitrate ($NH_4NO_3$) as electrolyte, respectively, in our previous publication [29]. Deionized (DI) water was used in all synthesis procedures.

*2.2 Synthesis of nitrogen doped reduced graphene oxide (NrEGO)*

A modified hydrothermal method was used to prepare NrEGO as illustrated in Fig. 1 [30]. 75 mg EGO was dispersed in 30 mL DI water and sonicated for 30 min. Then, 2.0 g urea was added into the dispersion. The mixture was sonicated for 3 h and then stirred for 1 h at room temperature. The mixture was transferred into 50 mL Teflon-lined steel autoclave to carry out hydrothermal treatment at 160 °C for 6 h. After which, the mixture was cooled down to room temperature and the products obtained were washed with DI water until pH=7. Finally, the material was dried at 60 °C in a vacuum oven and labeled as NrEGO.

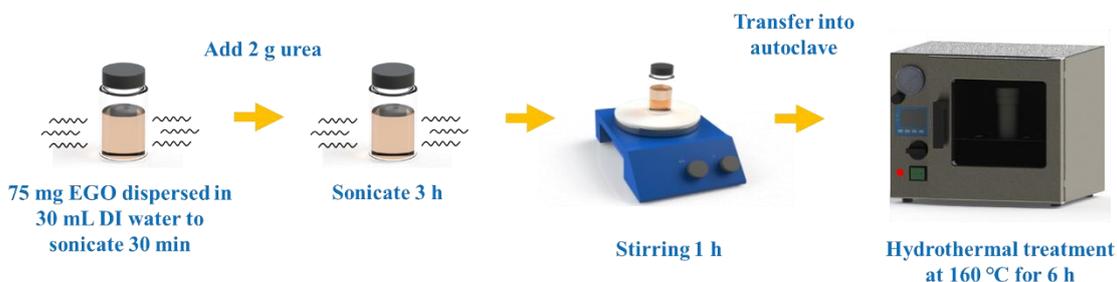

**Fig 1.** Synthesis routes of nitrogen-doped reduced electrochemically exfoliated graphene oxide (NrEGO).

*2.3 Synthesis of Pt/NrEGO$_x$-CB$_y$ catalysts*

A polyol reduction method developed in our group was used to synthesize the catalysts [29]. Firstly, different ratios (1:4, 2:3, 3:2, 4:1) of as-prepared NrEGO were mixed with CB (NrEGO$_x$-CB$_y$) to a total mass of 50 mg and dispersed into 5 mL DI water for 30 min with sonication. 200 mg of H$_2$PtCl$_6$·6H$_2$O dispersed in 20 mL ethylene glycol (EG) and added to the NrEGO$_x$-CB$_y$ dispersion. The mixture was sonicated for 2 h and then continuously stirred for another 1 h. Then, the mixture was transferred into 50 mL Teflon-lined steel autoclave and put in an oil bath at 120 °C to carry out the hydrothermal reduction process for 24 h, with continuous stirring. After cooling down to room temperature, the catalysts were washed with ethanol and DI water to remove chloride ions and organic residues. Finally, the Pt/NrEGO$_x$-CB$_y$ catalysts were dried at 60 °C in the vacuum oven and stored in ambient condition. To compare with the nitrogen doped support, none nitrogen doped hybrid catalysts (Pt/rEGO$_x$-CB$_y$) were synthesized by the same polyol reduction method with different ratios (1:4, 2:3, 3:2, 4:1) of rEGO and CB.

*2.4 Physical characterization*

EGO and NrEGO were characterized by ultraviolet-visible spectrophotometry (UV-vis) using a Shimadzu UV-2700 and Raman spectroscopy using a Renishaw inVia™ Confocal Raman microscope at 785 nm. The thermal stability of all EGO, NrEGO, NrEGO with CB hybrid support materials and as-prepared platinum catalysts were tested by thermogravimetric analysis (TGA) using TA instrument TGA 550 in an air atmosphere with the gas flow rate of 25 mL min$^{-1}$ and a temperature range of 25 °C - 900 °C increased by 20 °C min$^{-1}$. The X-ray diffraction (XRD) measurements were performed using a PANalytical X'Pert Pro X-ray diffractometer with CuK$_\alpha$ radiation ($\lambda_{K\alpha}$=0.15186 nm). The pattern was recorded in the *2θ* range of 10° to 100° with a step size of 0.016° for Pt/C and all synthesized Pt/NrEGO$_x$-CB$_y$. The average crystallite diameters, *d*, are calculated by Scherer equation: $d=k\lambda/\beta\cos\theta$ from Pt (111) and (220) only. *k* is shape factor with a typical value of 0.9, *λ* is the X-ray wavelength, *β* is the full width at half maximum (FWHM) in radians and *θ* is the Bragg angle. The chemical composition of the nitrogen-doped supports and catalysts was probed using X-ray photoelectron spectroscopy (XPS) on a Thermo NEXSA XPS operated using a mono-chromated Al kα X-ray source (1486.7 eV). All the data was treated with CasaXPS version 2.3.17PR1.1 (Casa Software LTD, Teignmouth, UK) and different materials were kept similar peak shape with almost same full width at half maximum (FWHM). The structure of the catalysts was characterized before and after in-situ

Accelerated Stress Tests (AST) by scanning electron microscopy (SEM) using a Quanta 650 SEM. The membrane electrode assembly was taken apart and a small amount of catalyst layer material was dispersed in acetone to prepare transmission electron microscopy (TEM) samples. The catalyst dispersion was dropped on holey carbon coated copper mesh TEM support grids so that the distribution of Pt nanoparticles on the supports and nanoparticle size could be investigated by a Thermo Fisher Titan scanning transmission electron microscopy (STEM) without those largest outlier particles. STEM was performed using a Thermo Fischer Titan ChemiSTEM equipped with a Cs probe corrector (CEOS) at a convergence angle of 21 mrad and a high angle annular dark field detector (HAADF) detector operated with inner angle of 55 mrad at 200 kV, probe current 180 pA.

*2.5 Electrochemical characterization*

Cyclic voltammetry (CV) was used to determine the specific electrochemical surface area (ECSA) of the catalysts in a three-electrode system using an Autolab PGSTAT30 potentiostat-galvanostat with an Autolab FRA2 module combined with a 10 A booster at room temperature. A mirror-polished glassy carbon (3 mm in diameter) with rotating disk electrode (RDE, Metrohm Autolab), platinum wire and Ag/AgCl (saturated KCl solution) were used as the working, counter and reference electrodes, respectively. The working electrode was prepared by mixing 5 mg catalyst with Nafion solution (5 wt%, 50 μL) and ethanol (950 μL) and sonicating for 30 min. Then a, 20 μL dispersion was dripped onto the working electrode and dried. CVs were tested under a potential range scanning from -0.2 to 1.0V with the scan rate of 50 mV s$^{-1}$ (N$_2$-saturated electrolyte, 1600 rpm) in a 0.5 M H$_2$SO$_4$ solution. ECSAs of catalysts were determined by integrating the hydrogen adsorption charges: [17,31]

$$\text{ECSAs} = Q_\text{h}/(Q_\text{m} \cdot \text{E.L}) \tag{1}$$

Where, $Q_\text{h}$ is the coulombic charge for hydrogen desorption (mC cm$^{-2}$). $Q_\text{m}$ is the adsorption charge for an atomically smooth surface, the value of $Q_\text{m}$ for Pt is 210 mC cm$^{-2}$. E.L represents the electrode Pt loading in g$_\text{Pt}$ cm$^{-2}$.

The ORR activity was measured by conducting linear sweep voltammetry (LSV), scanning from 1.0 to 0.0 V with a scan rate of 20 mV s$^{-1}$ under O$_2$-saturated 0.5 M H$_2$SO$_4$ solution at 400, 900, 1600, 2500 and 3600 rpm of rotating disk electrode (RDE). The mass-transport-corrected kinetic current was calculated by the Levich-Koutecký equation:

$$1/i = 1/i_\text{k} + 1/i_\text{d} = 1/B\omega^{1/2} + 1/i_\text{k} \tag{2}$$

$$B = 0.62nFC_0(D_0)^{2/3}v^{-(1/6)} \quad (3)$$

Where $i$, $i_k$, $i_d$ and $\omega$ represent the measured current density, the kinetic-limiting current density, the diffusion-limiting current density and RDE rotation rate, respectively. $B$ is the slope of the Levich-Koutecký curve. $n$ is the electron transfer number in ORR. $F$ is the Faraday constant (96485 C mol$^{-1}$), $C_0$ is the bulk concentration of O$_2$ (1.26×10$^{-3}$ mol L$^{-1}$), $D_0$ is the diffusion coefficient of O$_2$ (1.93×10$^{-5}$ cm$^2$ s$^{-1}$) in 0.5 M H$_2$SO$_4$, $v$ is the kinetic viscosity of water, 0.01 cm$^2$ s$^{-1}$ [32]. The onset potential ($E_0$) was defined as the potential at which the ORR starts, and the half-wave potential ($E_{1/2}$) was defined as the potential at which the ORR current is 50% of the limiting current density.

*2.6 Membrane electrode assembly (MEA) preparation*

To compare with our previous work in Pt/rEGO$_x$-CB$_y$ [29], the same carbon paper, microporous layer and 15% Nafion ionomer were used. The membrane electrode assembly consists of two electrodes (anode and cathode) and a polymer electrolyte. Both the anode and the cathode were made up of a gas diffusion layer (carbon paper TGPH 090, Toray), a micro-porous layer and a catalyst layer. The micro-porous layer was the same for each MEA and prepared by spraying the ink (90% ketjen black and 10% of 60 wt% of PTFE in 17 mL IPA) on top of carbon paper with the ketjen black loading of 1.0 mg cm$^{-2}$. Then, the catalyst layer was prepared by spraying catalyst ink (85% catalyst and 15% of 20 wt% Nafion solution in 15 mL acetone) over the micro-porous layer with Pt loading of 0.25 mg$_{Pt}$ cm$^{-2}$ at both anode and cathode. A 0.5 mg cm$^{-2}$ Nafion binding layer was sprayed on to the catalyst layer. The 1.5×1.5 cm$^2$ MEA containing two electrodes and a Nafion NR-212 membrane was prepared by hot-pressing at 135 °C for 4.5 min at 5.5 bar.

*2.7 Fuel cell performance test*

MEA performance was tested in a single cell at 40 °C, 50 °C and 60 °C with 100% relative humidity (RH), 100 mL min$^{-1}$ of both H$_2$ and O$_2$. Polarization curves and electrochemical impedance spectroscopy (EIS) were conducted using a Gamry potentiostat/galvanostat connected to the single cell. Three parallel experiments were carried out to test the polarization curves and evaluate the fuel cell performance. The kinetic parameters of Tafel slope were evaluated by fitting the polarization data to the semi-empirical equation in the lower current density of activation loss (< 0.10 A cm$^{-2}$): [33]

$$E_{\text{cell}} = E_0 - b\log(i) \quad (4)$$

Where $E_{\text{cell}}$ is the cell potential; $E_0$ is the open circuit potential, which is a constant depending on different catalyst cell at 60 °C with 100% RH; $b$ is the Tafel slope for oxygen reduction reaction in a single cell. The impedance spectra were recorded at a constant cell voltage of 0.60 V with the frequency swept from 10.00 kHz to 0.01 Hz at an AC signal amplitude of 100 mV and recorded 10 points per decade.

Two different Accelerated Stress Test (AST) protocols reported by Department of Energy (DOE) were carried out to evaluate the electrocatalyst degradation and carbon support corrosion [34]. All polarization curves were plotted by taking the average of three parallel experiments and a fresh MEA was used before each AST protocol. In the catalyst degradation AST, MEAs were operated at 60 °C with 100% RH $H_2$ (50 mL min$^{-1}$) and $N_2$ (100 mL min$^{-1}$) between 0.60 V and 1.00 V at a scan rate of 50 mV s$^{-1}$ for 30k cycles. Polarization curves and EIS were recorded before and after 30k cycles of AST at 60 °C with 100% RH $H_2$ (100 mL min$^{-1}$) and $O_2$ (100 mL min$^{-1}$). In the carbon support corrosion AST, another fresh MEAs were operated at 60 °C with 100% RH $H_2$ (50 mL min$^{-1}$) and $N_2$ (100 mL min$^{-1}$) between 1.00 V and 1.50 V at a scan rate of 500 mV s$^{-1}$. Polarization curves and EIS were also recorded before and after 5k cycles.

## 3. Results and discussion

*3.1 Physical characterization*

The reduction of EGO in the presence of urea can be directly observed by the color changes of the dispersion, from yellow brown to black in Fig. S1. This change was characterized by UV-Vis spectroscopy as shown in Fig. 2(a). An obvious absorption peak at 243 nm, which is the $\pi$-$\pi^*$ transition of aromatic C-C bonds in EGO, shifts to the right after using urea modification. This change indicates that the *p*-conjugated network within the graphene sheets was gradually restored and EGO has been reduced [35–37]. The shoulder at 300 nm in EGO, corresponding to n-$\pi^*$ transition of C=O bond, disappears after the hydrothermal treatment. This is also evidence that the oxygen functional groups of EGO have been reduced. Raman spectroscopy was used to characterize the quality of the graphene lattice and defects of NrEGO as shown in Fig. 2(b). The G peaks of EGO and NrEGO appeared at 1599.9 and 1595.9 cm$^{-1}$, respectively. The downshift of the G peaks by 4 cm$^{-1}$ from EGO to NrEGO is attributed to the restoration of the conjugated structure during the hydrothermal process [38]. It may also be related to the electron-donating capacity of the –$NH_2$ groups and graphitic N in n-type doping, as a similar downshift was observed in n-type doped graphite and carbon nanotubes (CNTs) [39–41]. Another characteristic feature of the NrEGO Raman spectrum is the intensity ratio of the D peak (1326.6 cm$^{-1}$) to the G peak ($I_D/I_G$), which represents the defects in the NrEGO flakes. It was found that $I_D/I_G$ of NrEGO slightly increases from 1.02 (EGO) to 1.11 (NrEGO). The increased value of $I_D/I_G$ indicates that NrEGO has more defects. The increased defects can be attributed to the introduction of N species [39,42].

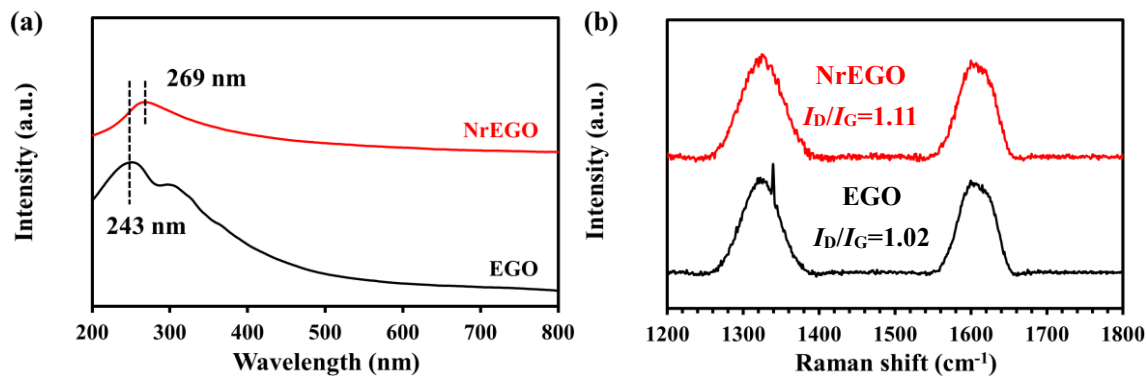

**Fig 2.** Characterization of EGO and NrEGO by (a) the whole wavelength analysis of UV-vis and (b) Raman spectra under 785 nm.

TGA curves were further used to investigate the thermal stability of EGO and NrEGO as shown in Fig. S2. Ignoring the weight loss of the materials due to water evaporation before 100 °C, the TGA curve for EGO displays two distinct stages of weight loss: the first stage is between 100 °C and 300 °C due to the decomposition of unstable oxygen functional groups, while the second stage (above 500 °C) belongs to the decomposition of carbon [43,44]. However, all N doped support materials do not show significant weight loss in the first stage, which also indicates that N doped graphene oxide has been reduced. Meanwhile, the carbon decomposition of all N doped support materials was observed after 700 °C, which are higher than that of in EGO. This indicates that the thermal stability of NrEGO and NrEGO$_x$-CB$_y$ are better than EGO [37].

X-ray photoelectron spectroscopy (XPS) was further used to confirm the extent to which functional groups have been reduced in NrEGO, as well as for quantifying different species of nitrogen groups and platinum. The C 1$s$ spectra of EGO in Fig. 3(a) shows three distinct components at 284.42 eV, 286.42 eV, 286.98 eV, which are assigned to C–C ($sp^2$), C–O and C=O [45]. However, after N was introduced into EGO, the C 1$s$ of NrEGO shows five distinctive components, including C–C $sp^2$ (284.48 eV), C–O (285.92 eV), C=O (287.94 eV), O–C=O (290.03 eV) and π-π* (291.90 eV) [20]. C–C $sp^2$ increases from 65.3% (EGO) to 79.8% (NrEGO) and the presence of the π-π* transition, due to localization of π electrons in the aromatic network [20]. The details are listed in Table S1. Meanwhile, the oxygen content decreases from 24.6% (EGO) to 4.0% (NrEGO), which agrees with the weight loss seen in TGA results.

Based on the high-resolution N 1*s* spectrum is shown in Fig. 3(b), where two species of N, *p*-phenyl (–NH$_2$, 399.32 eV) and graphitic N (402.21 eV), were identified on NrEGO [28,46]. A total of 2.4% N atoms were successfully introduced into graphene flakes, of which –NH$_2$ groups with 84.5% are the main N specie. The high content of –NH$_2$ can be explained with the reaction route shown in Scheme 1, the urea continues to release NH$_3$ at a relatively slow rate and the produced NH$_3$ reacts with oxygen functional groups (–OH and –COOH) on the surface of EGO to form –NH$_2$ [47]. After synthesizing Pt/NrEGO$_2$-CB$_3$, Fig. S3 shows an obvious extra N 1s peak at 398.76 eV, which can be assigned to a nitride, and indicates the presence of the Pt–N bond [48,49]. These –NH$_2$ groups and graphitic N belong to electron doping strategies (n type), which could improve effective carrier mobilities and adjust electrical conductivity [41]. –NH$_2$ groups increase electronic density to Fermi level and graphitic N facilitates 4e$^-$ electron transfer for ORR [50].

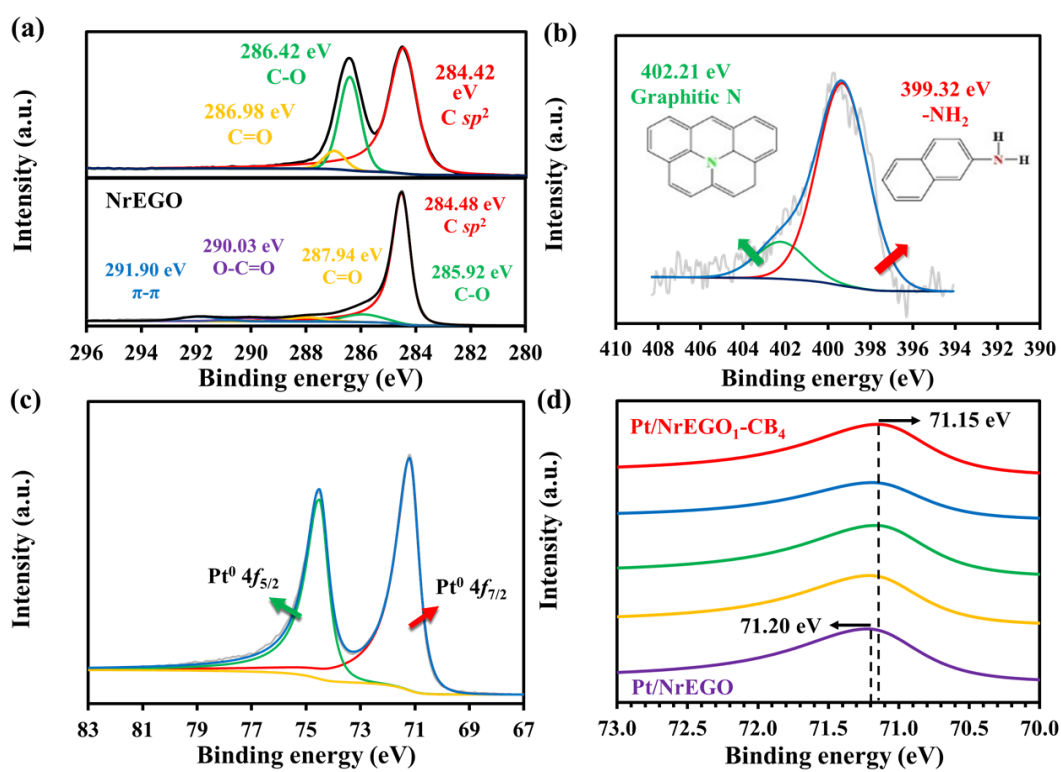

**Fig 3.** High-resolution XPS spectra of (a) C 1*s* and (b) N 1*s* in NrEGO, (c) N 1*s* and (d) Pt 4*f* in Pt/NrEGO$_2$-CB$_3$.

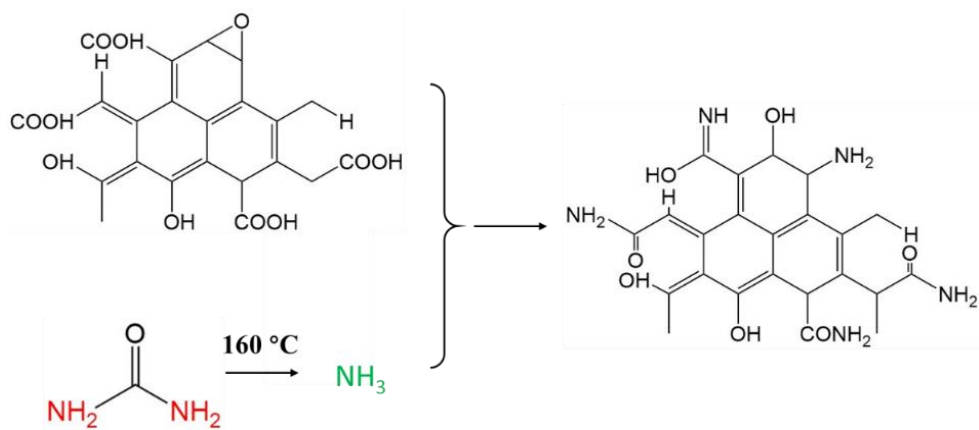

**Scheme 1.** A reaction route from the electrochemically exfoliated graphene oxide (EGO) to the nitrogen-doped reduced EGO (NrEGO).

The Pt $4f$ XPS spectra of Pt/NrEGO$_2$-CB$_3$ in Fig. 3(c) was deconvoluted into two peaks with binding energy values of 71.19 eV and 74.49 eV, corresponding to Pt$^0$ $4f_{7/2}$ and Pt$^0$ $4f_{5/2}$ respectively, showing that Pt is metallic. A positive shift of binding energy in Pt$^0$ $4f_{7/2}$ by 0.10 eV from 71.10 eV to 71.20 eV is seen with NrEGO amount increasing in Fig. 3(d). This can be attributed to both –NH$_2$ groups and graphitic N being always electro-negativity, which could enhance the interaction of positively charged Pt particles with N atoms. This is consistent with the findings of Xin et al and confirms the influence of –NH$_2$ groups and graphitic N on the electronic structure of Pt particles [28]. Xin et al also speculated that the electron density of Pt atoms delocalizing with the adjacent –NH$_2$ is greater than the formation of Pt–N bonds, which facilitates the formation and stability of metallic Pt [21,28]. Pt loading of the catalysts were investigated by means of ICP-OES and all catalysts exhibit around 50 wt% of Pt loaded on bi-support materials as shown in the supplementary information (Table S2). SEM and STEM characterizations were carried out to investigate the morphology of the as-prepared Pt nitrogen-doped bi-support catalysts. Fig. 4 shows SEM images of Pt/NrEGO$_x$-CB$_y$. Compared with pure Pt/C in Fig. 4(e), the agglomerated carbon spherical particles can be separated and dispersed to some extent, and it successfully intercalated into the interior of NrEGO flakes. The restacked NrEGO flakes were re-opened by the post-intercalation of CB spherical particles of Pt/NrEGO$_2$-CB$_3$ in Fig. 4(b). It is clearer that the restacked NrEGO flakes have more wrinkled edges after intercalation with carbon black. However, neither less NrEGO (NrEGO$_1$-CB$_4$ in Fig. 4(a)) nor more NrEGO flakes (NrEGO$_3$-CB$_2$ in Fig. 4(c) and NrEGO$_4$-CB$_1$ Fig. 4(d)) resulted in a good mixing of hybrid support materials.

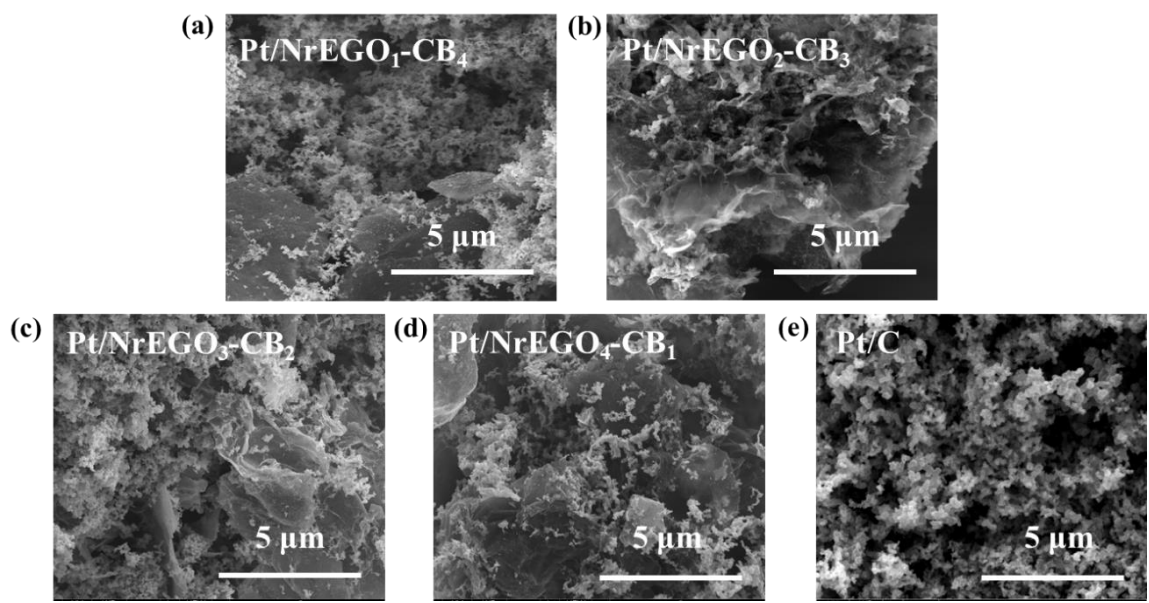

**Fig 4.** SEM images for the catalyst structure of (a) Pt/NrEGO$_1$-CB$_4$, (b) Pt/NrEGO$_2$-CB$_3$, (c) Pt/NrEGO$_3$-CB$_2$, (d) Pt/NrEGO$_4$-CB$_1$ and (e) commercial Pt/C.

The size distribution of PtNPs was analyzed by HAADF STEM images showing that PtNPs presents more uniform distribution on NrEGO$_2$-CB$_3$ in Fig 5(a). Statistical analysis of PtNPs on Pt/NrEGO$_2$-CB$_3$ gives a mean particle diameter of (4.88 ± 1.79) nm, and the mean size were determined (5.60 ± 1.80) nm in Pt/C and (5.70 ± 1.80) nm in Pt/rEGO$_2$-CB$_3$ in Fig. S4. Energy-dispersive X-ray spectroscopy (EDS) mapping data from Fig. 5(b) to Fig. 5(e) were employed to analyze the distribution of Pt, C and N elements in the Pt/NrEGO$_2$-CB$_3$. The high resolution HAADF STEM images in Fig. 5(f and g) show that PtNPs have a single-crystalline features and clear lattice fringes. The lattice spacing values of PtNPs are measured to be 2.24 Å and 1.80 Å, which are assigned to the expected (111) and (200) planes of Pt, respectively [28]. Meanwhile, The XRD patterns of all catalysts with different ratio of NrEGO and CB are displayed in the Fig. S5. The (111), (200), (220) and (311) peaks of Pt are at 39.9, 36.4, 67.7 and 81.4, and the (111) peak is the most significant, which agrees with STEM analysis. According to the Scherrer equation, the smallest Pt crystallite size corresponds to the Pt/NrEGO$_2$-CB$_3$ catalyst and is calculated to be (5.47 ± 0.40) nm, slightly smaller than (5.90 ± 0.44) nm in Pt/C and (5.86 ± 0.42) nm in no N-doped Pt/rEGO$_2$-CB$_3$. Crystallite size and particle size estimate from XRD and STEM are summarized in Table 1.

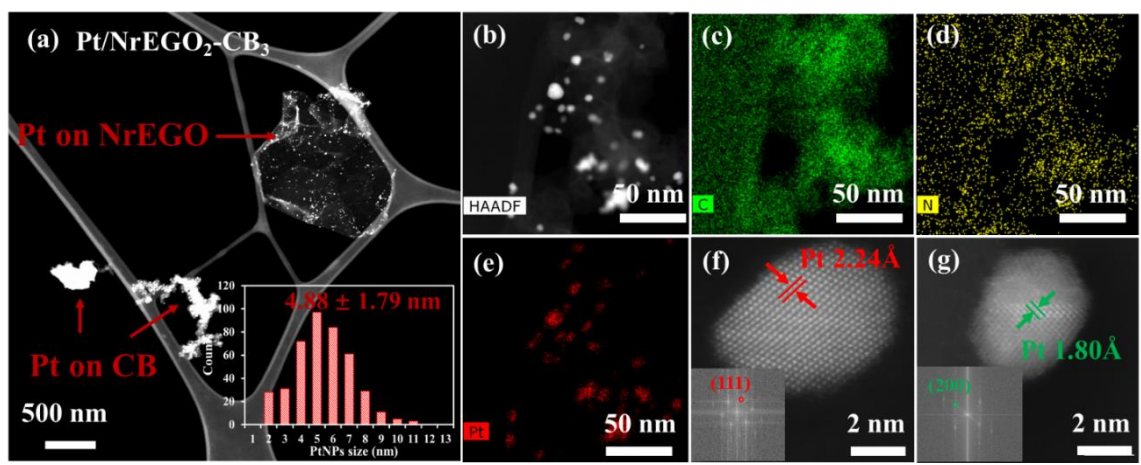

**Fig 5.** Typical HAADF STEM images of (a) Pt/NrEGO$_2$-CB$_3$ and the statistics on Pt particle size distribution by counting 421 particles. (b-e) HAADF STEM images of Pt/NrEGO$_2$-CB$_3$ and its corresponding EDS elemental maps of C, N, and Pt. (f and g) High resolution HAADF images of Pt/NrEGO$_2$-CB$_3$ showing lattice fringes of selected specific Pt particles with corresponding FFT patterns.

**Table 1.** Sherrer calculation by XRD results for Pt crystallite size and TEM results for Pt nanoparticles.

| Catalysts | FWHM | | Average crystallite size (nm) | Average nanoparticle size (nm) |
|---|---|---|---|---|
| | (111) | (200) | | |
| Pt/NrEGO$_2$-CB$_3$ | 1.44 | 1.71 | 5.47 ± 0.40 | 4.88 ± 1.79 |
| Pt/C | 1.33 | 1.58 | 5.90 ± 0.44 | 5.60 ± 1.80 |
| Pt/rEGO$_2$-CB$_3$ | 1.35 | 1.59 | 5.86 ± 0.42 | 5.70 ± 1.80 |

*3.2 Electrochemical characterization*

Cyclic voltammetry was used to evaluate the electrochemical surface area (ECSA) of all catalysts. Fig. 6(a) displays the hydrogen desorption peaks of Pt/NrEGO$_x$-CB$_y$ which had two types of electrodeposited H intermediates, underpotential-deposited hydrogen (UPD-H) and overpotential-deposited hydrogen (OPD-H) [51]. Among them, UPD-H (110) and (100) had a higher hydrogen binding energy (HBE), which is the dominating factor for the hydrogen oxidation reaction (HOR) activity. HOR contains a two-step reaction of the chemical adsorption step (Tafel reaction Eq. 5) and the electro-transfer step (Volmer reaction Eq. 6) [51–53]:

Tafel:
$$H_2 + 2^* \leftrightarrow 2H_{ad} \quad (5)$$

Volmer:

$$H_{ad} \leftrightarrow H^+ + e + * \tag{6}$$

After optimizing the ratio of NrEGO and CB, Pt/NrEGO$_2$-CB$_3$ achieves the maximum ECSA of 67 m$^2$ g$_{Pt}^{-1}$ with higher peak of (110) compared with that 51 m$^2$ g$_{Pt}^{-1}$ of Pt/C in Table 2. These ECSAs are also comparable with values of 19.11-168.20 m$^2$ g$_{Pt}^{-1}$ reported in recent publications concerning nitrogen doped catalysts as listed in Table S3 [54–59]. Meanwhile, Pt/NrEGO$_2$-CB$_3$ has a prominent platinum oxide reduction peak at 0.55 V in Fig. 6(a), while the same peak for Pt/C has a lower potential of only 0.47 V. This is attributed to the higher activity of smaller PtNPs resulting from the nitrogen incorporation. Linear sweep voltammetry was further used to analyze ORR activity. The onset potential ($E_0$), when ORR starts taking place, is a vital criterion for evaluating electrocatalytic activity. Fig. 6(b) shows the onset potentials measured by LSV are 0.829 for Pt/NrEGO$_1$-CB$_4$, 0.841 for Pt/NrEGO$_2$-CB$_3$, 0.861 for Pt/NrEGO$_3$-CB$_2$ and 0.844 V for Pt/NrEGO$_4$-CB$_1$, which present slightly higher onset potentials than that of Pt/C (0.824 V) and Pt/rEGO$_2$-CB$_3$ (0.819 V). With the amount of NrEGO flakes increasing, the voltage of the starting reaction gradually becomes higher, indicating that the presence of N facilitates the reaction to start. The half-wave potential ($E_{1/2}$) is 0.607 V for Pt/NrEGO$_2$-CB$_3$, which is almost same with Pt/NrEGO$_1$-CB$_4$ (0.609 V) and slightly lower than that of Pt/C (0.629 V) seen in Table 2. $E_{1/2}$ increases slightly to 0.622 V of Pt/NrEGO$_3$-CB$_2$, and then decreases quickly to 0.570 V of Pt/NrEGO$_4$-CB$_1$, indicating Pt/NrEGO$_4$-CB$_1$ having lower ORR activity. Pt/NrEGO$_2$-CB$_3$ has a lower diffusion-limiting current density of -6.169 mA cm$^{-2}$, which is similar with that of -5.556 mA cm$^{-2}$ for Pt/C and lower than that of -4.681 mA cm$^{-2}$ for Pt/rEGO$_2$-CB$_3$. Additionally, the kinetic current density derived from the different rotation rates (400-4800 rpm) of LSV measurements presented in Fig. S6 were used to calculate the electron transfer number ($n$) and the electrocatalytic activity in ORR. From the fitting results shown in Fig. 6(b), it was found that the electron transfer number for N-doped hybrid support catalysts increase from 3.8 to 4.2 as the NrEGO amount increases, suggesting an efficient 4e$^-$ reaction pathway with water as the product in ORR activity. This may be related to more –NH$_2$ groups and graphitic N content with NrEGO increase, which facilitates conductivity and electron transfer. The electron transfer number of Pt/NrEGO$_2$-CB$_3$ also has a relatively high value of ~4.1, which is higher than that of Pt/C (3.8).

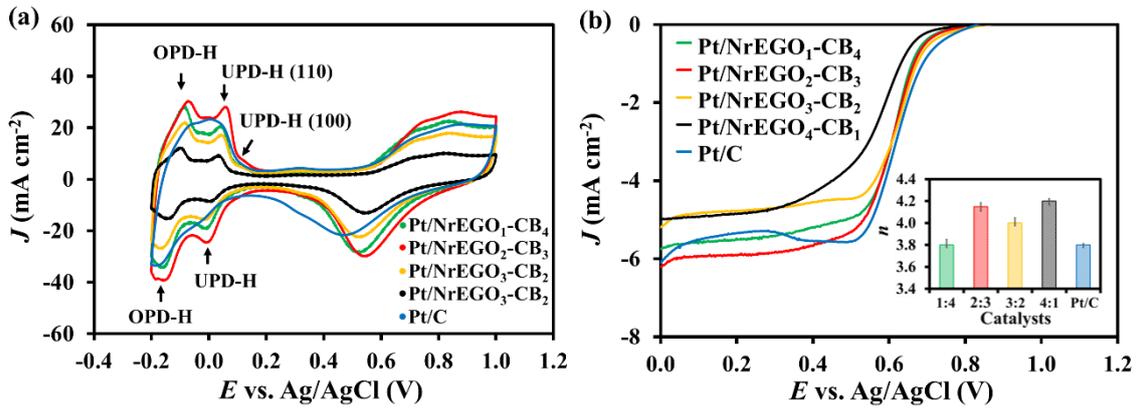

**Fig 6.** Electrochemical characterization of all catalysts: (a) ECSAs measured by cyclic voltammetry and (b) ORR activity and electron transfer number (inset) as measured by linear sweep voltammetry on all electrocatalysts.

**Table 2.** electrochemical characterization of all catalysts by cyclic voltammetry and linear sweep voltammetry.

| Catalysts | ECSA ($m^2\ g_{Pt}^{-1}$) | $E_{onset}$ (V) | $E_{1/2}$ (V) | $J_{limiting}$ (mA cm$^{-2}$) | $n$ |
|---|---|---|---|---|---|
| Pt/C | 51 | 0.824 | 0.629 | -5.556 | 3.8 |
| Pt/NrEGO$_1$-CB$_4$ | 56 | 0.829 | 0.609 | -5.793 | 3.8 |
| Pt/NrEGO$_2$-CB$_3$ | 67 | 0.841 | 0.607 | -6.169 | 4.1 |
| Pt/NrEGO$_3$-CB$_2$ | 41 | 0.861 | 0.622 | -5.170 | 4.0 |
| Pt/NrEGO$_4$-CB$_1$ | 25 | 0.844 | 0.570 | -4.978 | 4.2 |

*3.3 Fuel cell performance test*

Supplementary information Fig. S7 shows the polarization curves of Pt/NrEGO$_x$-CB$_y$ and Pt/NrEGO electrocatalysts at 40 °C, 50 °C and 60 °C with 100% RH of inlet gases. Performances of fuel cells with different catalysts increase when the temperature is increased and show the highest performance at 60 °C except Pt/NrEGO. In Fig. S7(e), the voltage of Pt/NrEGO decreases sharply which mostly because of the stacking graphene flakes in catalyst layer resulting in serious mass transport loss. Fig. 7(a) displays their comparison of the average polarization curves performance operating at 60 °C for all electrocatalysts. As control catalysts, commercial Pt/C and Pt/rEGO$_2$-CB$_3$ show maximum power densities of 0.710 W cm$^{-2}$ and 0.537 W cm$^{-2}$, respectively. However, Pt/NrEGO$_2$-CB$_3$ reached 1.18 W cm$^{-2}$ at the maximum current density of 2.22 A cm$^{-2}$ which can be detected in the instrument and achieved 4.748 W mg$_{Pt}^{-1}$. The power densities at 0.90 V and 0.60 V were used as indicators for the evaluation of the catalyst activity and MEA performance at operating voltage in practice. At 0.90 V,

Pt/NrEGO$_2$-CB$_3$ has the highest performance in all catalysts as shown in Fig. 7(b), which is related with its relatively lower Tafel slope (*b*) of 67 mV dec$^{-1}$ as shown in Table 3 and Fig. 7(c). It indicates a higher kinetic activity of the oxygen reduction reaction and a better electron transfer efficiency compared with that of 90 mV dec$^{-1}$ in commercial Pt/C. With the increase of NrEGO, Tafel slope increased to 128 mV dec$^{-1}$ for Pt/NrEGO$_3$-CB$_2$ and even 183 mV dec$^{-1}$ for Pt/NrEGO$_4$-CB$_1$.

At 0.60V, Pt/NrEGO$_1$-CB$_4$ (0.585 W cm$^{-2}$) and Pt/NrEGO$_2$-CB$_3$ (0.934 W cm$^{-2}$) electrocatalysts showed significant improvement in power density compared with no N-doped Pt/rEGO$_1$-CB$_4$ (0.267 W cm$^{-2}$) and Pt/rEGO$_2$-CB$_3$ (0.413 W cm$^{-2}$). The optimized Pt/NrEGO$_2$-CB$_3$ showed 1.55 times higher than commercial Pt/C (0.593 W cm$^{-2}$) at 0.60 V, as can be observed in Fig. 7(b). The power density of Pt/NrEGO$_2$-CB$_3$ showed 2.26 times higher than that of the optimized no nitrogen doped hybrid catalysts (Pt/rEGO$_2$-CB$_3$) in our previous publication as shown in Fig. S8 [29]. A higher content of NrEGO flakes produces the power density of 0.318 W cm$^{-2}$ in Pt/NrEGO$_3$-CB$_2$ with only a slight improvement compared with that of 0.269 W cm$^{-2}$ in Pt/rEGO$_3$-CB$_2$. If more NrEGO is added into the catalyst, instead of an increase of the power density, the effect obtained is the opposite and Pt/NrEGO$_4$-CB$_1$ provided a maximum power density of 0.130 W cm$^{-2}$ displaying lower performance than Pt/rEGO$_4$-CB$_1$. This is probably due to stacking of the high quality electrochemically exfoliated graphene oxide flakes because of the formation of π-π* transition confirmed by the XPS results during the hydrothermal reduction process and resulting in multi-layered graphite like structures. Therefore, both the surface area and anchoring sites of Pt are lost. On the other hand, the reduction process of synthetic catalysts exacerbates the stacking issues and prevents the carbon black intercalating sufficiently, which results in the difficulty of reactant gases (H$_2$ and O$_2$) approaching Pt active sites, thereby leading to a lower ORR activity. The polarization curves contain three important regions: activation loss (lower current density), ohmic loss (intermediate linear region) and mass transport loss (higher current density).

EIS was further used to quantify the resistance of each component inside the MEA in Fig. 7(d) and the specific values were displayed in Table 3. The equivalent circuit used for fitting EIS of Pt/C and Pt/NrEGO$_x$-CB$_y$ MEAs is constant phase shown in Fig. 7(e). In the high frequency of equivalent circuit model, the intercept at the real axis is related to the membrane resistance ($R_m$), representing the proton conductivity through the membrane [60]. Two distorted semi-circular arcs are presented in the Nyquist plots. The tiny arc at the high frequency (2-10 kHz) is defined as the interfacial resistance

($R_i$), which is influenced by the gap between catalyst layer and membrane [61]. The larger arc (0.1-2 kHz) is related to the charge resistance ($R_c$), which is attributed to the activity of cathode reaction [62,63]. The $R_m$ of all catalysts are almost the same because all MEAs use Nafion 212 membrane. However, $R_i$ increases with the amount of NrEGO because of the grain-boundary phenomena caused by multi phases components between the NrEGO flakes, carbon black and membrane, which lead to heterogeneities and more proton transport channels [64,65]. $R_c$ of Pt/NrEGO$_1$-CB$_4$ and Pt/NrEGO$_2$-CB$_3$ are the smallest and almost same with values of only 0.076 Ω cm$^2$ and 0.088 Ω cm$^2$, respectively. These $R_c$ values are much lower than that of 0.108 Ω cm$^2$ in the commercial Pt/C, 0.207 Ω cm$^2$ in Pt/NrEGO$_3$-CB$_2$ and 0.373 Ω cm$^2$ in Pt/NrEGO$_4$-CB$_1$. According to the SEM observation and ECSA results, this variation of $R_c$ occurs due to the lower ECSA caused by the loss of Pt active sites after stacking of NrEGO flakes. Stacking increases the difficulty of H$_2$ and O$_2$ reaching the surface of Pt, and thus, the reaction kinetics becomes slower [17].

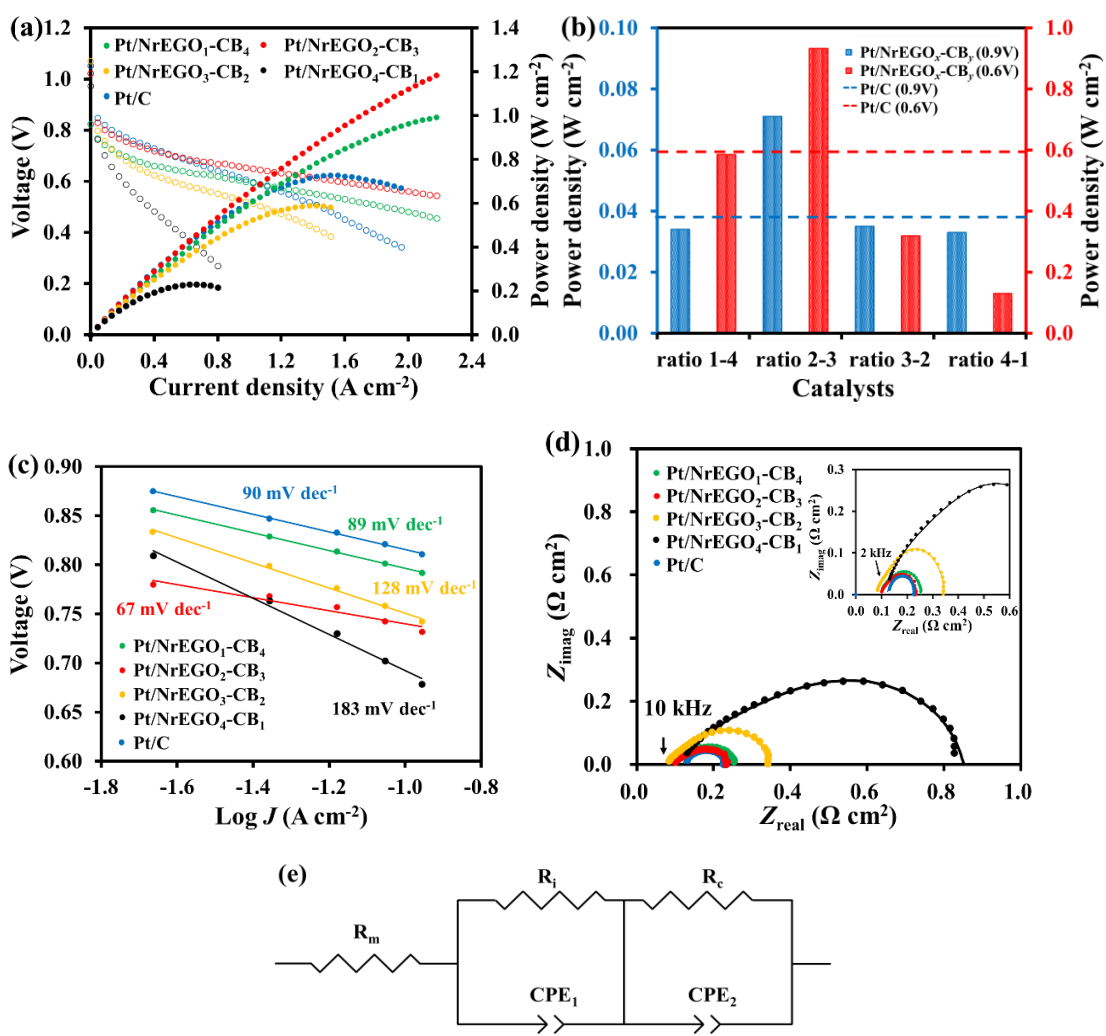

**Fig 7.** (a) Polarization curves of different ratio of N-doped hybrid support catalysts in MEAs and (b) comparation power density with same ratio of no N doped hybrid support catalysts (Pt/rEGO$_x$-CB$_y$) at 0.60 V. (c) Tafel slopes of Pt/NrEGO$_x$-CB$_y$ and Pt/C, respectively. (d) EIS of all catalysts and (e) the equivalent circuit diagram of a single cell.

**Table 3.** Comparation of electrochemical performance among N-doped hybrid support catalysts, commercial Pt/C and the optimized Pt/rEGO$_2$-CB$_3$.

| Catalysts | Power density at 0.60 V (W cm$^{-2}$) | Tafel slope $b$ (mV dec$^{-1}$) | $R$ (Ω cm$^2$) | | |
|---|---|---|---|---|---|
| | | | $R_m$ | $R_i$ | $R_c$ |
| Pt/C | 0.593 | 90 | 0.121 ± 0.004 | 0 | 0.108 ± 0.007 |
| Pt/NrEGO$_1$-CB$_4$ | 0.585 | 89 | 0.098 ± 0.009 | 0.080 ± 0.005 | 0.076 ± 0.011 |
| Pt/NrEGO$_2$-CB$_3$ | 0.934 | 67 | 0.101 ± 0.006 | 0.040 ± 0.002 | 0.088 ± 0.004 |
| Pt/NrEGO$_3$-CB$_2$ | 0.318 | 128 | 0.082 ± 0.016 | 0.056 ± 0.004 | 0.207 ± 0.003 |
| Pt/NrEGO$_4$-CB$_1$ | 0.130 | 183 | 0.107 ± 0.008 | 0.373 ± 0.011 | 0.373 ± 0.023 |

*3.4 Accelerated stress tests (ASTs)*

We also tested the durability of the optimized Pt/NrEGO$_2$-CB$_3$ and commercial Pt/C by accelerated stress tests (ASTs). To evaluate Pt catalyst degradation and corrosion resistance of catalyst supports, two different protocols of ASTs have also been carried out in single cells following the procedure described in the experimental section. The red curves in Fig. 8(a and d) display the polarization curves of Pt/C and Pt/NrEGO$_2$-CB$_3$, respectively, to evaluate the effect of catalyst degradation. After 30k low-potential cycles (0.60 V - 1.00 V), the polarization curve of commercial Pt/C deteriorated from 0.640 V to 0.548 V at 0.80 A cm$^{-2}$ as shown in Fig. 8(a), thus losing 92 mV. However, the loss observed in Pt/NrEGO$_2$-CB$_3$ is only 10 mV (from 0.659 V to 0.649 V), which is significantly lower than the commercial catalyst and achieves the target set by the DOE protocols (<30 mV at 0.8 A cm$^{-2}$ after 30k cycles) [34]. The Tafel slopes increased from 144 mV dec$^{-1}$ to 199 mV dec$^{-1}$ in Pt/C, which changed more than that of Pt/NrEGO$_2$-CB$_3$ from 124 mV dec$^{-1}$ to 132 mV dec$^{-1}$. It suggests that Pt/C activity decay and the electron transfer efficiency decreases much more quickly than that in Pt/NrEGO$_2$-CB$_3$. The results of ECSAs before and after ASTs presented in Fig. 8(b) show that Pt/C had a slight increase from 52 m$^2$ g$_{Pt}^{-1}$ to 56 m$^2$ g$_{Pt}^{-1}$ after 100 cycles for complete activation, and then, continuously decreased to 22 m$^2$ g$_{Pt}^{-1}$ over 30k cycles, a 58% ECSA loss, which is greater than that of 43% loss of Pt/NrEGO$_2$-CB$_3$ (from 63 m$^2$ g$_{Pt}^{-1}$ to 36 m$^2$ g$_{Pt}^{-1}$) in Fig. 8(e). This can be explained by EIS results obtained before and after AST in Fig. 8(c and f). $R_c$ in Pt/C increase from 0.089 Ω cm$^2$ to 0.165 Ω cm$^2$, which shows higher degradation than that of Pt/NrEGO$_2$-CB$_3$, increasing only from 0.088 Ω cm$^2$ to 0.151 Ω cm$^2$. The details are listed in Table S4. The lower increase of charge resistance can be attributed to the stable anchoring sites of Pt-N and C-N. Pt hopping directly over the N need an activation energy of 1.00 eV, which is much higher than that of 0.18 eV of Pt directly hopping over C atoms [21]. Therefore, Pt is not easy to migrate over the N atom, which results in the higher performance for Pt/NrEGO$_2$-CB$_3$ after AST.

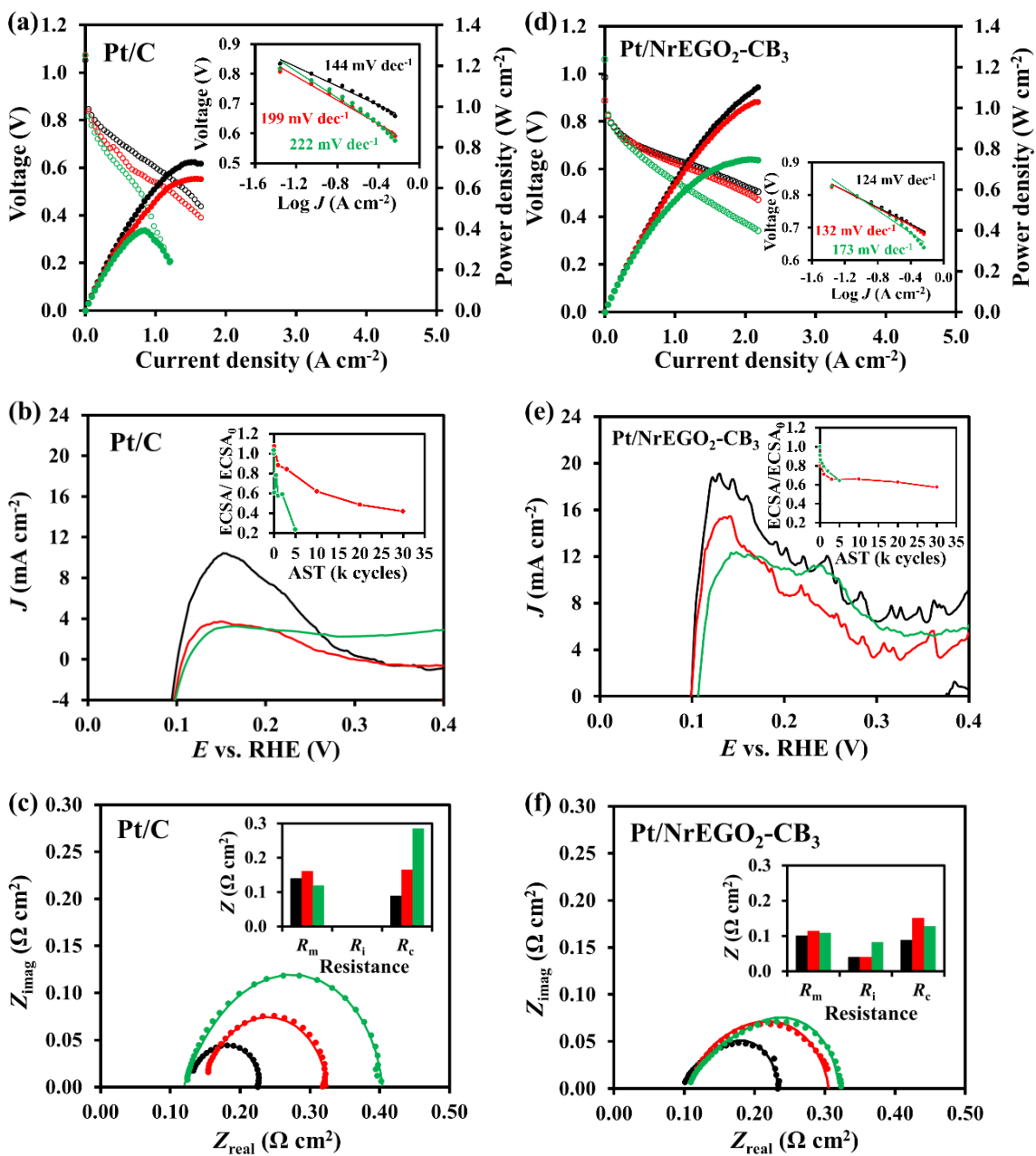

**Fig 8.** Electrochemical performance of catalysts before and after catalyst degradation and carbon corrosion ASTs. The measurements of Pt/C in (a) polarization curves with Tafel slopes, (b) ECSA and (c) EIS before and after ASTs. The measurement of Pt/NrEGO$_2$-CB$_3$ in (d) polarization curves with Tafel slops, (e) ECSA and (f) EIS before and after ASTs. • fresh catalysts. • after 30k cycles catalysts degradation ASTs. • after 5k cycles support corrosion ASTs.

The green curves in Fig. 8(a and d) show the polarization curves of Pt/C and Pt/NrEGO$_2$-CB$_3$ after carbon corrosion test. The maximum power density of Pt/C significantly decreases from 0.727 W cm$^{-2}$ to 0.393 W cm$^{-2}$, keeping 64% of initial performance after 5k cycles. It is not able to reach 1.5 A cm$^{-2}$ and the Tafel slope increases to 222 mV dec$^{-1}$. However, Pt/NrEGO$_2$-CB$_3$ reduces from 0.891 W cm$^{-2}$ to 0.692 W cm$^{-2}$, keeping 78% at 1.5 A cm$^{-2}$. The Tafel slope of Pt/NrEGO$_2$-CB$_3$ is 173 mV dec$^{-1}$, which is also lower than that of Pt/C. These indicate that the carbon in Pt/C is poorly resistant to corrosion and this has a huge influence on the activity of the Pt nanoparticles. This is also demonstrated by ECSA and EIS. The rapid reduction of ECSA by 76% in Fig. 8(b) and the significant charge resistance of Pt/C increasing to 0.285 Ω cm$^2$ after ASTs in Fig. 8(c). Meanwhile, the ECSA of Pt/NrEGO$_2$-CB$_3$ decreases by only 36% in Fig. 8(e) and $R_c$ increases slightly to 0.128 Ω cm$^2$ in Fig. 8f. These demonstrate that NrEGO$_2$-CB$_3$ as a supporting material has higher resistance to corrosion. According to the mechanism of carbon oxidation reaction (carbon corrosion, COR), the effect of catalyst performance goes through the following steps: the formation of $C_{(s)}^+$ during electrochemical process (7), the formation of carbon oxides by hydration (8 and 9), carbon monoxide poisoning of Pt (10 and 11): [66–68]

$$C \rightarrow C_{(s)}^+ + e^- \tag{7}$$

$$C_{(s)}^+ + H_2O \rightarrow CO_{surf} + 2H^+ + e^- \tag{8}$$

$$CO_{surf} + H_2O \rightarrow CO_2 + 2H^+ + 2e^- \tag{9}$$

$$CO_{surf} + Pt \rightarrow Pt - CO_{ads} \tag{10}$$

$$Pt - CO_{ads} + H_2O \rightarrow CO_2 + 2H^+ + 2e^- \tag{11}$$

Therefore, carbon oxidation reaction (COR) process results in dynamic process including leaching, Pt agglomeration/aggregation with migration and coalescence with electrochemical process through Ostwald ripening [17,69,70]. The bigger Pt particles are easily detached with carbon corrosion and carried out with the discharged water. Subsequently, Pt or Pt clusters will be poisoned by the produced CO during the support corrosion process. The defects created by –NH$_2$ groups and graphitic N is more stable to prevent carbon corrosion [71].

## 4. Conclusions

In conclusion, *p*-phenyl (–NH$_2$) groups and graphitic N doped reduced electrochemically exfoliated graphene oxide intercalated by carbon black as a highly durable Pt catalyst hybrid support for the hydrogen fuel cell was prepared through a facile hydrothermal method. The 2.4% of nitrogen atoms introduced into reduced electrochemically exfoliated graphene oxide achieved the platinum nanoparticles size at (4.88 ± 1.79) nm with ECSA up to 67 m$^2$ g$_{Pt}^{-1}$.

It was found through testing in the low-temperature hydrogen fuel cell that reactant gases (H$_2$ and O$_2$) transfer problems due to stacking of graphene flakes are prevented by optimising the ratio of NrEGO to CB. The optimized Pt/NrEGO$_2$-CB$_3$ MEA could achieve high Pt efficiency of 4.748 W mg$_{Pt}^{-1}$, which is 2 times higher than commercial Pt/C MEA. A higher power density performance of 0.934 W cm$^{-2}$ at an operating potential of 0.60 V in the low-temperature hydrogen fuel cell. This performance is 1.55 times higher than that of commercial Pt/C.

In terms of the durability, Pt/NrEGO$_2$-CB$_3$ MEA has shown excellent performance in both Pt degradation and support corrosion. After 30k cycles of accelerated Pt degradation test, Pt/NrEGO$_2$-CB$_3$ MEA showed only a slighter decay of 10 mV compared with 92 mV in Pt/C MEA at 0.80 A cm$^{-2}$, which could meet the target of 30 mV reported by Department of Energy. Pt/NrEGO$_2$-CB$_3$ had only 43% loss of ECSAs and the charge resistance ($R_c$) of Pt/NrEGO$_2$-CB$_3$ had only changing from 0.089 Ω cm$^2$ to 0.165 Ω cm$^2$. On the other hand, after 5k cycles of accelerated carbon corrosion test, Pt/C MEA only maintained 54% performance and could not even reach 1.5 A cm$^{-2}$. However, Pt/NrEGO$_2$-CB$_3$ MEA kept 78% performance and showed 0.692 W cm$^{-2}$ at 1.5 A cm$^{-2}$, indicating high resistance to corrosion. The ECSA of Pt/NrEGO$_2$-CB$_3$ decreases only 36% and its charge resistance increase only to 0.128 Ω cm$^2$, which is less than that of Pt/C. These enhanced performance and durability of Pt/NrEGO$_2$-CB$_3$ were attributed to the function of N as a bridge between Pt and C, namely: i) –NH$_2$ groups and graphitic N has stronger interaction to anchor Pt particles, which decrease Pt catalyst degradation. ii) –NH$_2$ groups and graphitic N doped electrochemical exfoliation graphene oxide and carbon black (N-C) was more stable, which improved corrosion resistance than a single carbon material.

**Acknowledgements**

This work was supported by the Engineering and Physical Sciences Research Council (EPSRC) EP/P009050/1 and EP/S021531/1. This paper acknowledges the Henry Royce Institute for Advanced Materials, funded through the EPSRC grants EP/R00661X/1, EP/S019367/1, EP/P025021/1 and EP/P025498/1. SJH acknowledges funding from the European Commission H2020 ERC Starter grant EvoluTEM (715502).

# Graphical Abstract

Platinum supported on nitrogen doped reduced electrochemically exfoliated graphene oxide/carbon black hybrid support shows excellent performance with more stable platinum activity and higher resistance to carbon corrosion than that of commercial Pt/C.

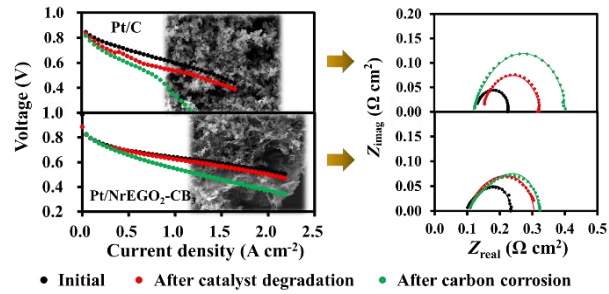

**Doped graphene/carbon black hybrid catalyst giving enhanced oxygen reduction reaction activity with high resistance to corrosion in proton exchange membrane fuel cells**

Zhaoqi Ji [1, 2], Jianuo Chen [2], María Pérez-Page [2*], Zunmin Guo [2], Ziyu Zhao [2], Rongsheng Cai [3], Maxwell T. P. Rigby [3], Sarah J. Haigh [3], Stuart M. Homes [2*].

[1] School of Automotive Engineering, Harbin Institute of Technology, Weihai, 264209, Shandong, China

[2] Department of Chemical Engineering and Analytical Science, The University of Manchester, Manchester, M13 9PL, U.K.

[3] Department of Materials, The University of Manchester, Manchester, M13 9PL, U.K.

* Corresponding authors: stuart.holmes@manchester.ac.uk (S. M. Holmes); maria.perez-page@manchester.ac.uk (M. Pérez-Page)

**Table S1** Content of C 1$s$, N 1$s$ spectra in different hybrid supports.

| Samples | O content (%) | N 1s | C 1s | | | | |
|---|---|---|---|---|---|---|---|
| | | | C $sp^2$ | C-O | C=O | O–C=O | π-π* |
| EGO | 24.6 | 0.5 | 65.3 | 29.0 | 5.7 | - | - |
| NrEGO$_1$-CB$_4$ | 4.2 | 0.5 | 84.5 | 9.4 | 1.4 | 3.5 | 1.2 |
| NrEGO$_2$-CB$_3$ | 4.0 | 0.8 | 83.2 | 9.6 | 2.4 | 3.7 | 1.1 |
| NrEGO$_3$-CB$_2$ | 5.4 | 1.4 | 81.0 | 11.2 | 1.9 | 3.4 | 2.5 |
| NrEGO$_4$-CB$_1$ | 4.3 | 1.9 | 80.6 | 10.6 | 2.4 | 3.8 | 2.6 |
| NrEGO | 4.0 | 2.4 | 79.8 | 10.5 | 2.5 | 3.1 | 4.1 |

**Table S2** ICP-OES analysis for Pt loading on the catalysts.

| Catalysts | Pt loading (wt%) |
|---|---|
| Pt/NrEGO$_1$-CB$_4$ | 50.62 ± 1.80 |
| Pt/NrEGO$_2$-CB$_3$ | 48.69 ± 1.65 |
| Pt/NrEGO$_3$-CB$_2$ | 48.51 ± 2.20 |
| Pt/NrEGO$_4$-CB$_1$ | 47.07 ± 3.41 |

**Table S3** Literature review for the electrochemical properties and hydrogen fuel cell performance of

N-doped carbon catalysts.

| Catalysts | ECSA (m$^2$ g$_{Pt}^{-1}$) | N content (wt%) | $E_{1/2}$ (V) | Power density (W cm$^{-2}$) | Ref. |
|---|---|---|---|---|---|
| Pt/CNTs | 55.60 | | | ~0.600 | [1] |
| Pt@NC/C | 98.30 | | | | [2] |
| Pt/NCB | 168.20 | 0.80 | | | [3] |
| Pt/NCA | 19.11 | | | | [4] |
| Pt/NOMC | | 2.20 | 0.680 | | [5] |
| Pt/G-NCNTs-1/2 | 118.40 | 4.67 | | | [6] |
| Pt/NCNT | 24.90 | 7.50 | ~0.600 | | [7] |
| Pt/NSWCNH | 83.00 | 1.00 | 0.904 | | [8] |
| Pt/ED-CNT | 67.50 | 4.74 | 0.790 | 1.040 | [9] |
| N-CNT/Pt/NC | 75.50 | 9.90 | ~0.800 | 0.676 | [10] |
| Pt/NCNC$_{700}$ | 60.40 | 27.60 | 0.838 | | [11] |
| Pt/CPPA$_{800}$ | 31.00 | 10.50 | ~0.800 | | [12] |
| Pt/N-CQD/G | 123.40 | 2.40 | 0.880 | | [13] |
| Pt@N-graphene foam | 84.90 | 7.17 | 0.895 | 0.394 | [14] |
| This work | 67.00 | 2.40 | 0.607 | > 1.187 | |

**Table S4** Resistance changes before and after two AST protocols (catalyst degradation and carbon corrosion).

| AST protocols | Pt/C | | | Pt/NrEGO$_2$-CB$_3$ | | |
|---|---|---|---|---|---|---|
| | $R_m$ | $R_i$ | $R_c$ | $R_m$ | $R_i$ | $R_c$ |
| Initial | 0.140 | 0 | 0.089 | 0.101 | 0.040 | 0.088 |
| After 30k catalyst degradation | 0.161 | 0 | 0.165 | 0.114 | 0.040 | 0.151 |
| After 5k carbon corrosion | 0.119 | 0 | 0.285 | 0.109 | 0.083 | 0.128 |

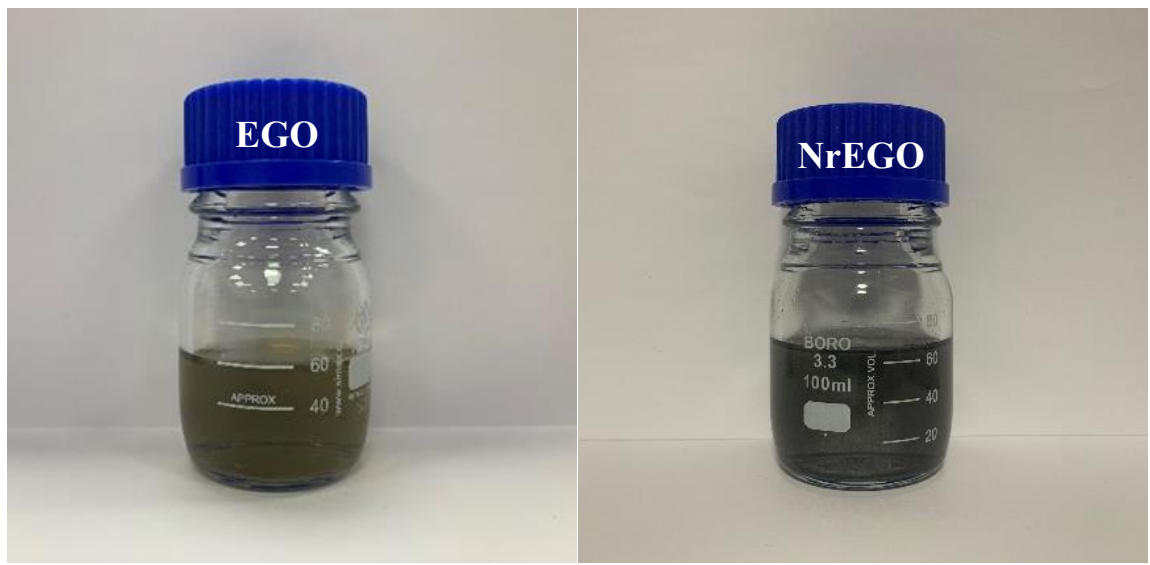

**Fig. S1** Color changes of EGO and NrEGO.

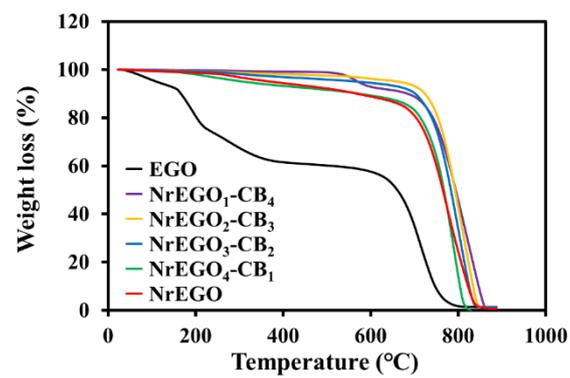

**Fig. S2** TGA analysis for the thermal stability of all N-doped catalysts support.

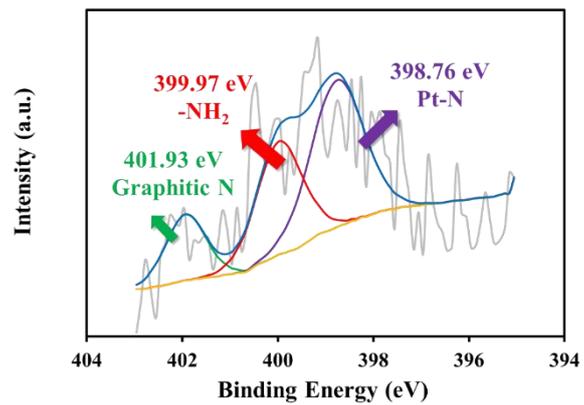

**Fig. S3** High-resolution XPS spectra of N 1s in Pt/NrEGO$_2$-CB$_3$.

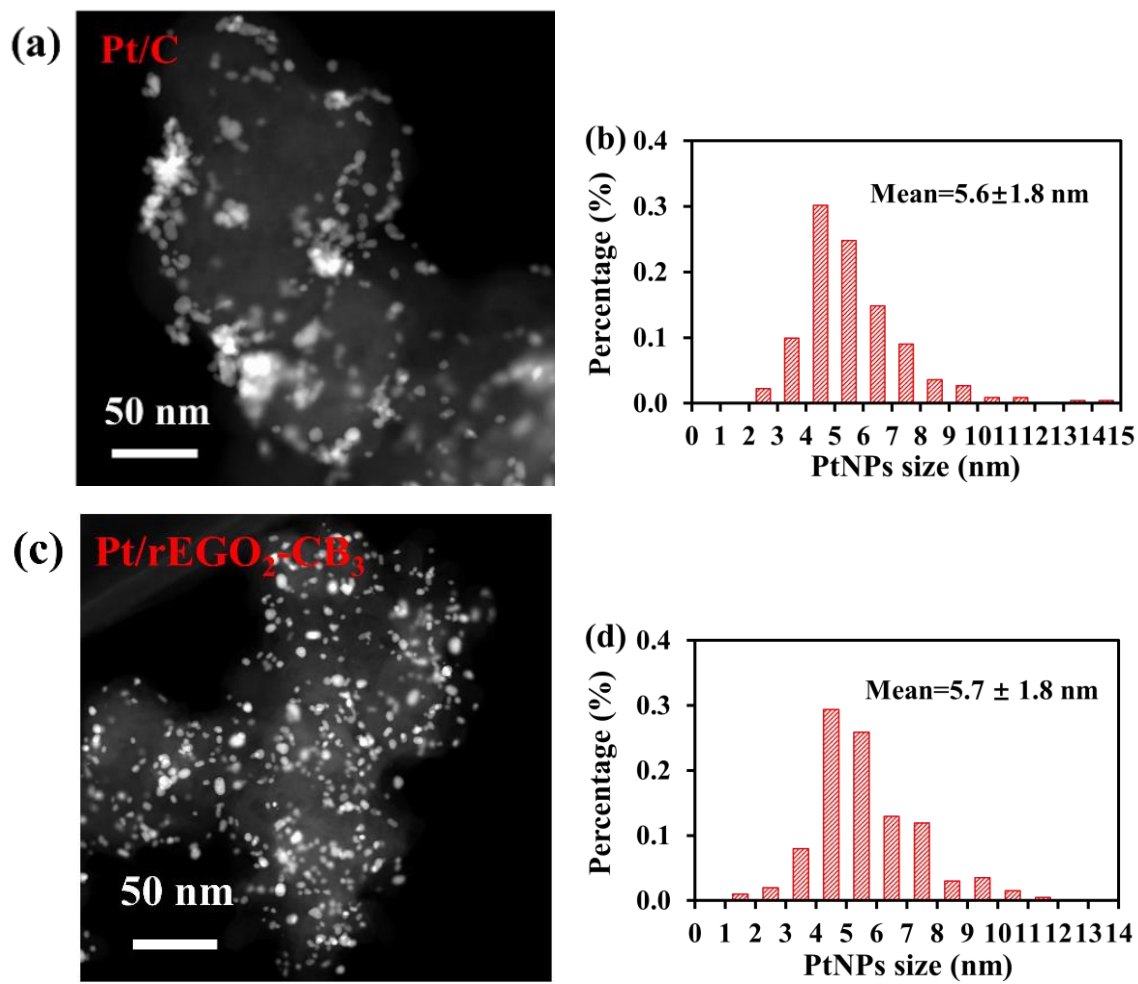

**Fig. S4** Typical HAADF STEM images of (a) Pt/C and (c) Pt/rEGO$_2$-CB$_3$. Statistical analysis of PtNPs size distribution of (b) Pt/C by counting 222 particles and (d) Pt/rEGO$_2$-CB$_3$ by counting 201 particles.

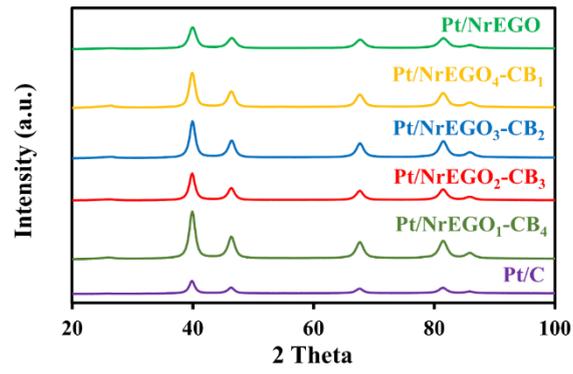

**Fig. S5** The XRD spectra comparing Pt/C with different ratios of hybrid catalyst support Pt/NrEGO$_x$-CB$_y$ and Pt/NrEGO.

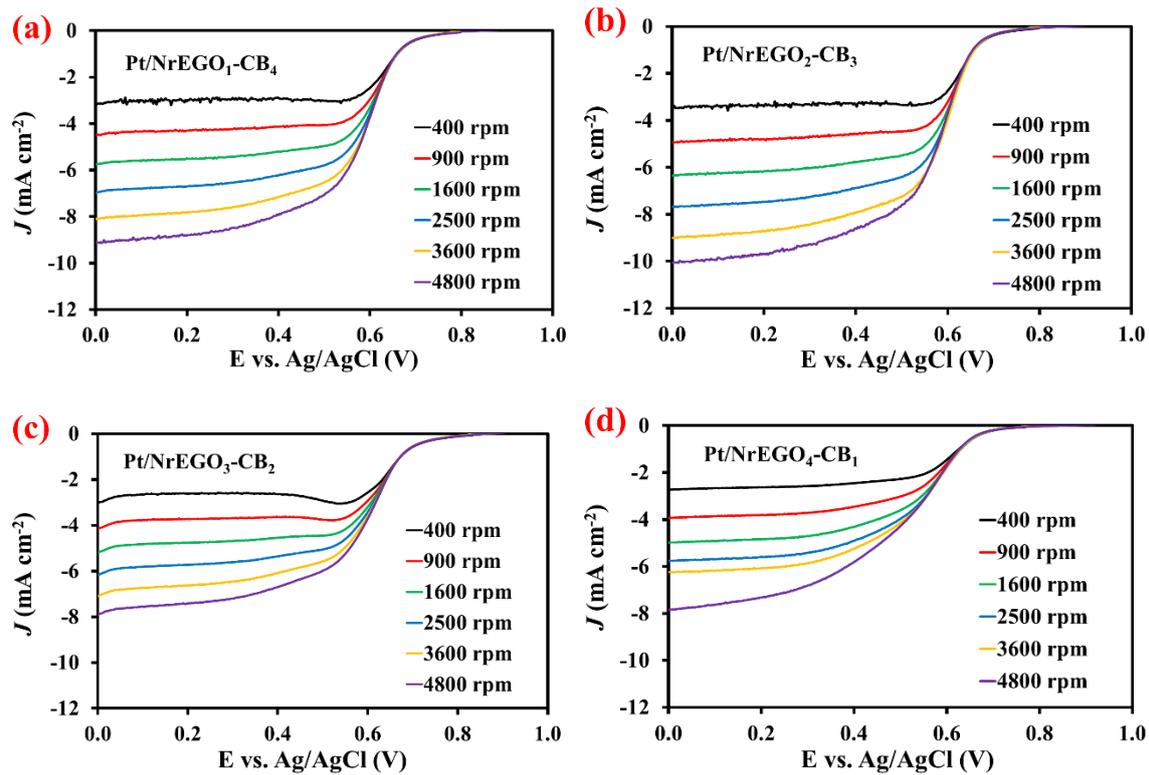

**Fig. S6** Linear sweep voltammetry of all nitrogen doped catalysts at different rotation speed for the evaluation of ORR activity.

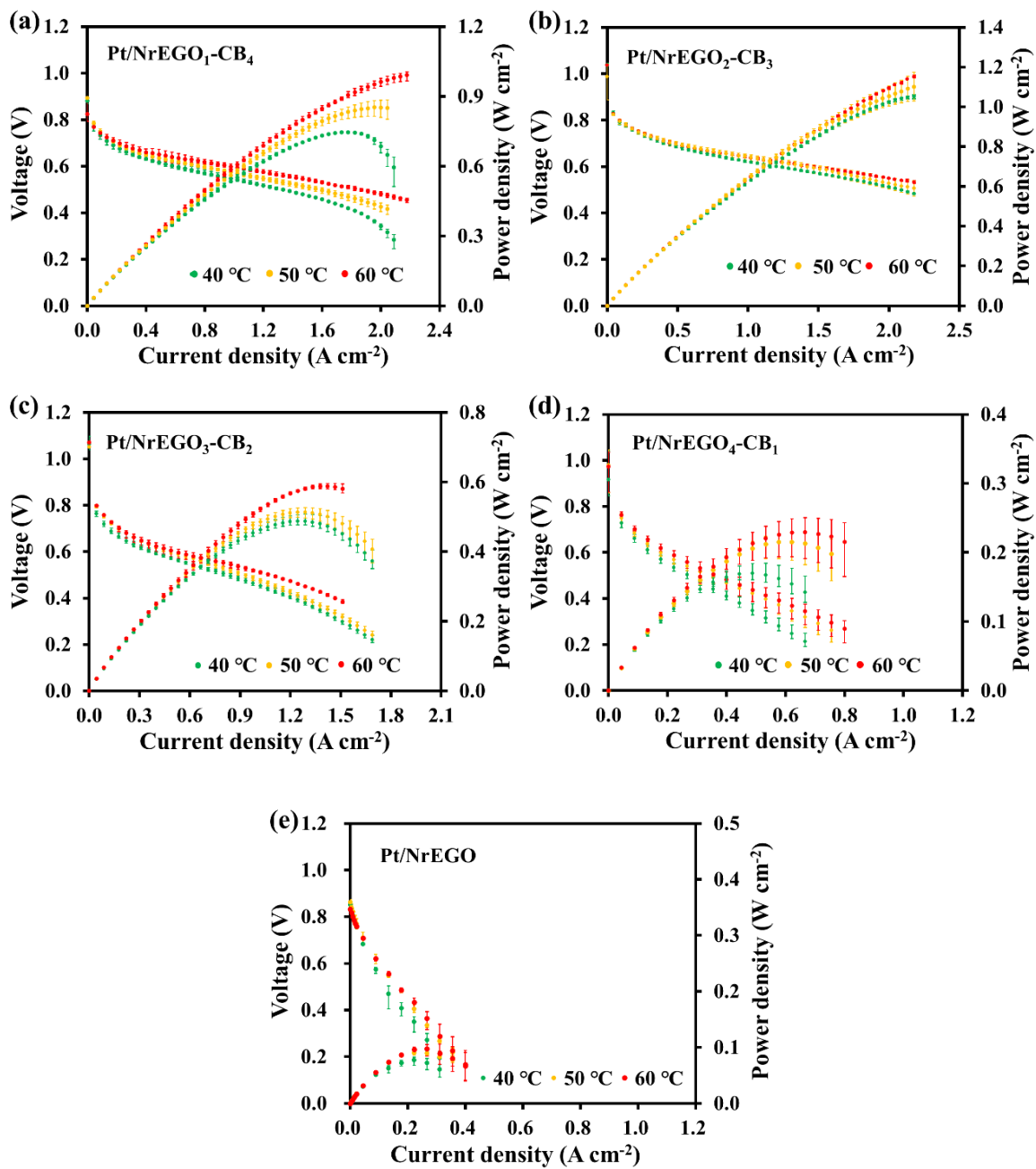

**Fig. S7** All nitrogen doped catalyst performance at 40 °C, 50 °C and 60 °C.

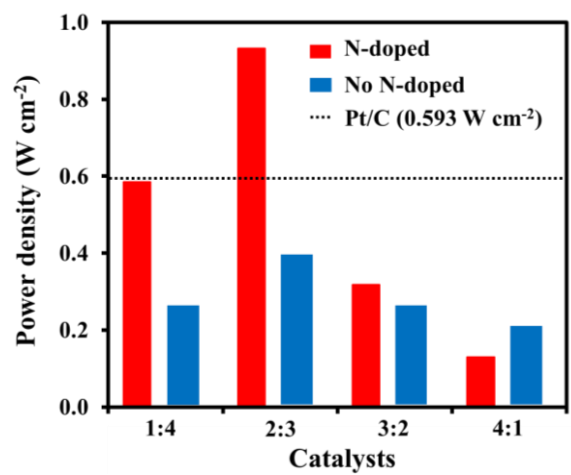

**Fig. S8** The comparation of Pt/NrEGO$_x$-CB$_y$ and Pt/rEGO$_x$-CB$_y$ performance at 0.60 V.